\documentclass[11pt,a4paper,superscriptaddress,prb,amsmath,amssymb,aps]{revtex4}
\usepackage{graphicx}
\usepackage{bm}
\usepackage[all]{xy}

\newcommand{\qed}{\nobreak \ifvmode \relax \else
      \ifdim\lastskip<1.5em \hskip-\lastskip
      \hskip1.5em plus0em minus0.5em \fi \nobreak
      \vrule height0.75em width0.5em depth0.25em\fi}

\begin{document}
\title{Time-dependent corrections to effective rate and event statistics in Michaelis-Menten kinetics}

\author{N.~A. Sinitsyn} \email{nsinitsyn@lanl.gov}
\affiliation{Theoretical Division, Los Alamos National Laboratory, Los
  Alamos, NM 87545, USA}
\affiliation{New Mexico Consortium, Los
  Alamos, NM 87545, USA}

\author{Ilya Nemenman} \email{ilya.nemenman@emory.edu} \affiliation{Department of Physics, Department of Biology, and Computational and
  Life Sciences Strategic Initiative, Emory University, Atlanta, GA
  30322, USA}

\pacs{03.65.Vf, 05.10.Gg, 05.40.Ca}

\begin{abstract} 
  We generalize the concept of the geometric phase in stochastic
  kinetics to a noncyclic evolution. Its application is demonstrated
  on kinetics of the Michaelis-Menten reaction. It is shown that the
  nonperiodic geometric phase is responsible for the correction to the
  Michaelis-Menten law when parameters, such as a substrate
  concentration, are changing with time. We apply these ideas to a
  model of chemical reactions in a bacterial culture of a growing
  size, where the geometric correction qualitatively changes the
  outcome of the reaction
  kinetics. \end{abstract}

\date{\today}

\maketitle

\section{Introduction}

Biochemical reactions are typically characterized in stationary {\em
  in vitro} environments with the hope that their measured properties
will hold {\em in vivo}. There are clearly many important
physiological reasons why this extrapolation may fail. In this
article, we focus on one particular reason that has little to do with
the physiology, but rather derives from the fact that rates of complex
chemical reactions may have non-trivial corrections due to slow,
adiabatic drift of (internal) kinetic parameters of the system \cite{astumian-09jcc}.

The class of phenomena we study is related to the celebrated Berry's
phase in driven quantum mechanical systems \cite{berry-84}, which
predicted a contribution to the phase of an adiabatically changing
wave function in the form of a integral over the parameter trajectory.
Since the original Berry's discovery a number of its generalizations
were proposed, e.g., to nonabelian and nonadiabatic regimes. Similar
geometric phases were also found in other fields, for example, in
dissipative dynamics
\cite{landsberg-92,kagan-91,sinitsyn-08prb,sinitsyn-08jpa}.

Recently, new geometric phases where studied in the domain of purely
classical stochastic kinetics
\cite{sinitsyn-07epl,sinitsyn-07prl,sinitsyn-07prb,Parrondo98,jarzynski-08prl,horowitz-09jstat}. They were shown
to be responsible for the stochastic pump and other ratchet-like
effects, and thus they are of clear importance for the theory of
chemical enzymes, and specifically molecular motors operating in
strongly stochastic environment \cite{shi,astumian-pnas}. This finding
raises possibilities of various generalizations of the geometric
phase. For example, recently its nonadiabatic counterpart was
introduced in Ref.~\cite{ohkubo-08}, and it was shown to be
responsible for a non-adiabatic current contribution that has no
analog under stationary conditions.
 
In this Letter, we study another generalization of the geometric phase
in stochastic kinetics, namely to a nonperiodic evolution in the
parameter space. While its quantum and optical versions were explored
in a series of studies
\cite{pati-98,jain-98,noncyclic1,noncyclic2,noncyclic3,noncyclic4,noncyclic5,noncyclic6,noncyclic7},
their role is still largely unclear. In this work, we show that the
gauge invariant noncyclic geometric phase in stochastic kinetics can
be unambiguously defined, and that it can be naturally interpreted as
being responsible for the leading nonadiabatic correction in the
expression for stochastic fluxes, which can {\em qualitatively} change
kinetics of a chemical reaction.

\section{Generating function for the Michaelis-Menten reaction}

The Michaelis-Menten (MM) reaction \cite{MM} is the most fundamental
and the simplest enzymatic biochemical process. It describes a
catalytic conversion of one type of molecules, called the substrate,
into another type, called the product, via an intermediate reaction
with an enzyme. Schematically the MM reaction can be represented as
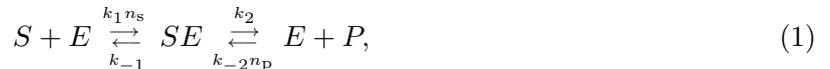
\begin{equation}
  S +E^{\,\,{k_1 n_{\rm s} \atop \longrightarrow}}_{\,\,\,\,  {{\longleftarrow}\atop k_{-1}}}\,\,
 SE^{\,\,\,\,\,\,{k_2 \atop \longrightarrow}}_{\,\,{\longleftarrow
     \atop k_{-2}n_{\rm p}}}\,E+P,
\label{MM}
\end{equation}
where $S$ and $P$ denote substrate and product respectively, $s$ and
$p$ stand for their concentrations, and $E$ is the enzyme molecule.
$S$ and $P$ interact via creating a complex $SE$ which is unstable and
dissociates either back into $E$ and $S$ or forward into $E$ and $P$.
In the simplest version of the MM mechanism, enzymes catalyze the
process but are not modified in any reactions. However,
generalizations are certainly possible \cite{english-etal-06}. 

In their 1913 article \cite{MM}, Michaelis and Menten considered a
strongly nonequilibrium situation, neglecting the backwards $E+P$
association, which can be done for $n_{\rm p}\ll k_1n_{\rm s}/k_{-2}$. However,
here we keep this reaction for generality. If the number of $S$ and
$P$ molecules is much larger than that of the enzymes, the latter have
to perform many substrate conversions each in order to change $S$ and
$P$ concentrations noticeably. This is traditionally used to simplify
the reaction kinetics since one can assert that enzymes operate in a
quasi steady state at current substrate and product concentrations.

Stochastic kinetics of the conversion of $S$ into $P$ is conveniently
described by the moments generating function $Z(\chi,t)$ (mgf) and the
cumulants generating function $S(\chi,t)$ (cgf) defined as
\cite{sinitsyn-07epl,szabo-06,nazarovFCS}
\begin{equation}
Z(\chi,t)=e^{S(\chi,t)}=
\sum_{n=-\infty}^{\infty} P_{n}e^{in\chi},
\label{pgf1}
\end{equation}
where $P_{n}$ is the probability to find net $n$ product molecules
generated during the observation time $t$ (back conversion is counted
with the negative sign). For a small number of enzymes, they can be
considered statistically independent over short periods of time, and
the cgfs are additive. Thus we will restrict our study only to the
case of a single enzyme without  loss of generality. 

It is convenient to introduce additional generating functions
$U_E=\sum_{n=-\infty}^{\infty} P_{nE}e^{in\chi}$ and
$U_{SE}=\sum_{n=-\infty}^{\infty} P_{nSE}e^{in\chi}$, where $P_{nE}$
and $P_{nSE}$ are the probabilities that, at a given time, the net
number of generated product molecules is $n$, and the enzyme is in the
unbound/bound state. Then the master equation for the entire process
is \begin{equation} \begin{array}{l}
    \frac{d}{dt} P_{nE} = -(k_1n_{\rm s}+k_{-2}n_{\rm p})P_{nE} +k_{-1}P_{nSE}+k_2P_{(n-1)SE},\\
    \frac{d}{dt} P_{nSE} = -(k_{-1}+k_{2})P_{nSE}
    +k_{1}n_{\rm s}P_{nE}+k_{-2}n_{\rm p}P_{(n+1)E}. \end{array} \label{master-2}
\end{equation} Multiplying (\ref{master-2}) by $e^{i\chi n}$ and
summing over $n$ we find the equation for the generating functions:
\begin{equation}
  \frac{d}{dt}\left(\begin{array}{l}
 U_{E}\\ 
 U_{SE} 
\end{array}\right)=-\hat{H}(\chi,t) \left(\begin{array}{l}
 U_{E}\\ 
 U_{SE} 
\end{array}\right),
\label{master-3}
\end{equation}
where
\begin{equation}
\hat{H}(\chi,t)=\left(
\begin{array}{cc}
k_1n_{\rm s} + k_{-2}n_{\rm p}   & -k_{-1}-k_2 e^{i \chi} \\
-k_1n_{\rm s}-k_{-2}n_{\rm p}e^{-i\chi}   & k_{-1}+k_2
\end{array} \right).
\label{hchi}
\end{equation}
If we set $n=0$ at initial moment $t=0$, then the initial conditions for (\ref{master-3})
are $U_{E}(t=0)=p_E(0)$, and $U_{SE}(t=0)=p_{SE}(0)$, where $p_E(0)$ and $p_{SE}(0)$
 are probabilities that the enzyme is free/bound, respectively. Additionally, note that $Z(\chi,t)=U_{E}(\chi,t)+U_{SE}(\chi,t)$.   
Thus the formal solution for the mgf (\ref{pgf1}) can be expressed as an average of the evolution operator
\begin{equation}
Z(\chi,t)= \langle 1\vert \hat{T}\left(e^{-\int_{0}^{t}\hat{H}(\chi,t) dt}\right) \vert  p(0) \rangle,
\label{pdf2}
\end{equation}
where $\langle 1 \vert =(1,1)$, $\vert
p(0)\rangle=(p_{E}(0),p_{SE}(0))^T$, and $\hat{T}$ is the
time-ordering operator.

Before we proceed with the case where parameters are time dependent,
it is instructive to look first at the stationary regime. To simplify
(\ref{pdf2}), one can find normalized left and right eigenvectors
$\langle u_{0/1} \vert$, $\vert u_{0/1} \rangle$ and corresponding
eigenvalues $\epsilon_{0/1}$ of the operator $\hat{H} (\chi)$,
where indices $0$ and $1$ correspond to the two eigenvalues with the
smallest and the largest real parts, respectively. There is one left
and one right eigenvectors for each eigenvalue.

Every vector, such as $\vert p(0)\rangle$ can be expressed as a sum of eigenvectors of $\hat{H} (\chi)$, for example,
\begin{equation}
\vert p(0) \rangle = \langle u_{0}\vert p(0) \rangle \vert u_{0} \rangle + \langle u_{1}\vert p(0) \rangle \vert u_{1} \rangle,
\label{ppp}
\end{equation}
where we define $\langle \alpha \vert \beta \rangle = \alpha_1\beta_1+\alpha_2 \beta_2$ to be a standard scalar product of two vectors.
Substituting (\ref{ppp}) into (\ref{pdf2}), for the time-independent Hamiltonian we find the steady state mgf,
\begin{equation}
Z_{\rm st}(\chi,t)=  e^{-\epsilon_0(\chi)t+\ln\left( \langle 1\vert u_0 \rangle\langle u_{0}\vert p(0)\rangle\right)} +
 e^{-\epsilon_1(\chi)t+\ln \left(  \langle 1\vert u_1 \rangle\langle u_{1}\vert p(0) \rangle\right)} ,
\label{pdf4}
\end{equation}
At time scales $t\gg \max[1/k_{-1},1/k_2,1/(k_1n_{\rm
  s}),1/(k_{-2}n_{\rm p})]$,
the second term in (\ref{pdf4}) is exponentially suppressed in
comparison to the first, and the expression for the mgf
simplifies to
\begin{equation}
Z_{\rm st}(\chi,t) \approx  e^{-\epsilon_0(\chi)t+\ln\left( \langle 1\vert u_0 \rangle\langle u_{0}\vert p(0)\rangle\right)}  .
\label{pdf5}
\end{equation}
Terms analogous to $-\epsilon_0(\chi)t$ in (\ref{pdf5}) have been
studied previously \cite{nazarovFCS,sinitsyn-07epl}. The second term
is less threaded: this is the boundary term that does not grow with
time and depends on the initial conditions and the averaging over the
final states of the enzyme. One can disregard it in comparison to the
first contribution when $t\to\infty$. However, we note that its
relative effect decays as $1/t$, i.e., not exponentially. We will keep
the boundary term in the following discussion because it will play an
important role to restore the gauge invariance of the nonperiodic
geometric phase.

At the first look, the boundary term leads to a contradictory result
after setting $t\rightarrow 0$, i.e at the initial moment of the
evolution. In this limit, the boundary term does not disappear, namely
\begin{equation}
\left.S_{\rm bnd}\right|_{t=0}=\ln \left( \langle 1 \vert u_0(0) \rangle \langle u_0(0) \vert p(0) \rangle \right) \ne 0 . 
\label{sgeom2}
\end{equation}
However, we expect $\left.S_{\rm bnd}\right|_{t=0}$ to be zero, since
$\left.n\right|_{t=0}=0$, so the mgf should be identically equal to
unity. The apparent contradiction is resolved by noting that
(\ref{pdf5}) was derived assuming $t\to\infty$, and it is simply an
invalid approximation for $t=0$. In other words, the boundary term is
responsible for the initial fast relaxation to the stationary regime.
For more insight, one can calculate the contribution of the boundary
term to the average number of generated product molecules. Using the
normalization condition $p_{SE}(0)=1-p_{E}(0)$ one can find
\begin{equation}
n_{\rm  bnd}=-i\left.\frac{\partial \left.S_{\rm bnd}\right|_{t=0}}{\partial \chi} \right|_{\chi=0} = 
\frac{(k_2 + k_{-2}n_{\rm p})(k_2 + k_{-1} - Kp_{E}(0))}{K^2},
\label{bound1}
\end{equation}
where $K=k_{-1}+k_2+k_1n_{\rm s}+k_{-2}n_{\rm p}$. If one assumes that the initial
probability $p_E(0)$ for the enzyme to be free is at the equilibrium
value $p_E(0)=(k_2+k_{-1})/K$, then (\ref{bound1}) produces $
n_{\rm bnd}=0$, as expected. To confirm this, one can also derive
(\ref{bound1}) by a standard master equation approach. That is,
calculating the average number of new product molecules $n_{\rm bnd}(t)$, one would find that, after a sufficiently long time,
\begin{equation}
n(t)=n_{\rm bnd}+\frac{k_1k_2n_{\rm s}-k_{-1}k_{-2}n_{\rm p}}{K}t.
\label{avcharge}
\end{equation}
The second term in Eq. (\ref{avcharge}) is the average number of the
product molecules produced during time $t$ at a steady state. It is
the standard prediction of the reversible MM theory, and the first
term is a correction, which is nonzero when the initial state of
enzymes is not the same as its steady state.

\section{Noncyclic geometric phase in stochastic kinetics} 
Assume now that there are several slowly time-dependent parameters in
the model. We will group them in a vector $\mathbf \lambda$. In the case
of the MM process, one can view these time-dependent parameters as
concentrations of the substrate and the product, ${\mathbf
  \lambda}=(n_{\rm s},n_{\rm p})$. However, the discussion in this section is
completely general.
 
Following Ref.~\cite{sinitsyn-07epl} we partition the time into small
intervals, over which kinetic rates can be considered almost constant.
We insert the resolution of the identity operator, $\hat{1}=\vert
u_0(t) \rangle \langle u_0(t)\vert+\vert u_1(t) \rangle \langle
u_1(t)\vert$, in (\ref{pdf2}) after every such an interval. One can
find then that the boundary term becomes $S_{\rm bnd}=\ln \left( \langle
  1 \vert u_0(t) \rangle \langle u_0(0) \vert p(0) \rangle \right)$.
Importantly, it is no longer gauge invariant, i.e., it is sensitive to
the redefinition of eigenstates of the Hamiltonian (\ref{hchi}) such
as $|u_0\rangle \rightarrow e^{\phi({\mathbf \lambda})}|u_0 \rangle$ and
$\langle u_0| \rightarrow \langle u_0 | e^{-\phi({\mathbf \lambda})}$.
Therefore, taken alone, it has no direct physical meaning. 

It will be convenient to rewrite the boundary term as a sum of a gauge
invariant part and a term that is an integral from a pure derivative,
i.e., \begin{equation} S_{\rm bnd}=\left. S_{\rm bnd}\right|_{t=0}+\int_{{\bf c}} {\bf
    P}\cdot d {\mathbf \lambda},\quad {\bf P} =\partial_{{\mathbf \lambda}}
  \ln \langle 1 \vert u_{0} \rangle, \label{sbound} \end{equation}
where ${\bf c}$ is the contour in the space of the variable parameters. By
analogy with Ref.~\cite{sinitsyn-07epl}, and including the boundary
contribution (\ref{sbound}), the mgf in the
quasi steady state limit can be written as an exponent of a sum of two
terms,
\begin{equation} Z(\chi)=e^{S_{\rm geom}(\chi)+S_{qst}(\chi)},
  \label{mgf3} \end{equation} where \begin{equation}
  S_{\rm qst}(\chi)=-\int_{0}^{t} \epsilon_0 (\chi,t') dt'+\left.S_{\rm bnd}\right|_{t=0}
  \label{sqst3} \end{equation} is the quasistationary part of the
generating function averaged over time. This is the part that
morphs into the steady state result (\ref{pdf5}) for fixed values
of all parameters. 

The other term in (\ref{mgf3}), \begin{equation}
  S_{\rm geom}=\int_{\bf c} [{\bf P}({\mathbf \lambda}) -{\bf A}({\mathbf
    \lambda})] \cdot d{\mathbf \lambda},\quad {\bf A}({\mathbf
    \lambda})=\langle u_0 \vert \partial_{{\mathbf \lambda}} u_0 \rangle,
  \label{sgeom} \end{equation}
is the {\em geometric phase} contribution responsible for additional
reaction events. ${\bf A}$ is called the {\em Berry connection}.
$S_{\rm geom}$ has no analog in the strict steady state regime.

Note that, unlike in Ref.~\cite{sinitsyn-07epl}, we do not assume a
periodic evolution of parameters. Therefore, the term involving the
integral of the Berry connection ${\bf A}$ over the path in the
parameter space, $-\int_{\bf c} {\bf A}({\mathbf \lambda}) \cdot d{\mathbf
  \lambda}$, is generally not gauge invariant. However, one can easily
check that the non-gauge-invariant contribution due to the boundary
term exactly cancels the non-gauge-invariant part of the contour
integral from ${\bf A}$.
 
We further mention that the definition (\ref{sgeom}) differs somewhat
from those used for the non-cyclic geometric phase in quantum
mechanics. For example, Refs.~\cite{pati-98,jain-98} define the
noncyclic geometric phase as $\gamma_{\rm gp}=\int_{\bf c} [ {\bf
  A}({\mathbf \lambda}) -{\bf P}({\mathbf \lambda}) ]\cdot d{\mathbf \lambda}$,
where ${\bf P} =
-{\rm Im} \left( \frac{\langle
      u({\mathbf \lambda}(0))\vert
    \partial_{{\mathbf \lambda}}u({\mathbf \lambda}) \rangle } {\langle u({\mathbf
      \lambda}(0))\vert u({\mathbf \lambda}) \rangle} \right)$. In the
present context, the meaning of such definition is unclear,
while the geometric phase defined in (\ref{sgeom}) is derived directly
from the exact representation of the mgf.

Since ${\bf P}$ is a pure gauge, it is important only when looking at
an evolution along an open path in the parameter space. If the
parameter vector ${\mathbf \lambda}$ returns to its initial value at the
end of the evolution, the expression (\ref{sgeom}) becomes equivalent
to the full-period geometric phase defined in
Ref.~\cite{sinitsyn-07epl}.

\section{Corrections to Michaelis-Menten law}

Consider now the average product creation rate in the MM system under
the slow parameter evolution. The average number of new product
molecules is $\langle n (t)\rangle = -i\left(
  \partial Z(\chi,t)/\partial \chi \right)_{\chi=0}$. Therefore, just
like the full cgf, the average rate of the product production $\langle
J \rangle =d\langle n (t)\rangle/dt$ can be written as a sum
of the quasistationary $J_{\rm qst}$ and the geometric $J_{\rm geom}$
contributions \begin{equation} \langle J \rangle = J_{\rm geom}+J_{\rm
    qst}=\frac{d}{dt} \left.\frac{\partial S_{\rm geom}}{\partial
      \chi} \right|_{\chi=0} + \left.\frac{\partial \epsilon_0
      (\chi,t)}{\partial \chi} \right|_{\chi=0}. \label{curcur}
\end{equation} The geometric phase is time-dependent only via the
time-dependence of the parameter vector ${\mathbf \lambda}$. In the
case of MM reaction with time-dependent concentrations $n_{\rm s}$ and
$n_{\rm p}$, the time derivative of the first term in (\ref{curcur})
can be expressed as $d/dt \rightarrow (dn_{\rm s}/dt)\partial/\partial
n_{\rm s}+(dn_{\rm s}/dt)\partial/\partial n_{\rm p}$. Substituting the eigenvectors and
eigenvalues of $\hat{H}(\chi,{\bf \lambda})$ into (\ref{curcur}), we find \begin{eqnarray}
  J_{\rm qst}&=&\frac{(k_1n_{\rm s}(t))k_2-(k_{-2}n_{\rm p}(t))k_{-1}}{K},
  \label{curcur2} \\
J_{\rm geom}&=&-(k_2+k_{-1})
\frac{(k_2+k_{-2}n_{\rm p}(t))(k_1\dot{n}_{\rm
    s}(t)+k_{-2}\dot{n}_{\rm p}(t))}{K^3}.
\label{curcur3}
\end{eqnarray}   
One can recognize $J_{\rm qst}$ as the average current for a steady
state with fixed values of parameters. In fact, (\ref{curcur2}) is
what is known as the Michaelis-Menten law. However, our results show
that this law is not exact when concentrations of the substrate and the
product have their own time-dependent evolution. The geometric
contribution is the first correction to the Michaelis-Menten kinetics
that becomes nonzero when the substrate/product concentrations change with
time. Specifically, in the most frequent case $n_{\rm p}\approx 0$, the
average rate of the coarse grained MM reaction per one enzyme becomes
\begin{equation}
\langle J \rangle \approx \frac{k_2n_{\rm s}}{n_{\rm s} +\frac{k_2 +k_{-1}}{k_1}}-(k_2+k_{-1})
\frac{k_2k_1\dot{n}_{\rm s}(t)}{(k_1n_{\rm s}+k_2+k_{-1})^3}.
\label{rater}
\end{equation}
That is, even in this case, the time-dependence of the substrate
concentration introduces corrections to the reaction rate.

It is possible to understand the result (\ref{curcur3}) with a simpler
approach, which, unfortunately, is hard to generalize for higher
current cumulants to demonstrate the geometric nature of the effect
for all of them. The probability $p_E$ of the enzyme to be unbound
evolves according to the master equation \begin{equation} \frac{d}{dt}
  p_E =
  -[k_1n_{\rm s}(t)+k_{-2}n_{\rm p}(t)]p_E + (k_2+k_{-1})(1-p_E), \label{dpe}
\end{equation} with the solution \begin{equation} p_E(t)
  =(k_2+k_{-1})\int_{0}^t e^{ - \int_{t_1}^{t}
    [k_1n_{\rm s}(\tau)+k_{-2}n_p(\tau)+k_2+k_{-1}]d\tau} dt_1. \label{dpe1}
\end{equation} The lower limit in this integral is not important
because we work in the adiabatic approximation, which means that the
information about the initial state is quickly forgotten. The exponent of
the integral over $\tau$ in (\ref{dpe1}) is then
\begin{multline} e^{ - \int_{t_1}^{t}
    [k_1n_{\rm s}(\tau)+k_{-2}n_{\rm p}(\tau)+k_2+k_{-1}]d\tau} \\\approx
  e^{-[k_1n_{\rm s}(t)+k_{-2}n_{\rm p}(t)+k_2+k_{-1}](t-t_1)}\left( 1+
    \frac{k_1\dot{n}_{\rm s}(t)+k_{-2}\dot{n}_{\rm p}(t)}{2} (t-t_1)^2 \right).
  \label{apr} \end{multline} 
Performing the remaining integration we
find the expression for the probability of the enzyme to be unbound:
\begin{equation} p_E \approx
  \frac{k_2+k_{-1}}{K}+\frac{(k_2+k_{-1})(k_1\dot{n}_{\rm
      s}(t)+k_{-2}\dot{n}_{\rm p}(t))}
  {K^3}. \label{probe} \end{equation} From (\ref{probe}), one can
calculate the average reaction rate and check that indeed, it is the
sum of the quasi-stationary and the geometric components determined in
(\ref{curcur2}) and (\ref{curcur3}), \begin{equation}
  J(t)=(1-p_E(t))k_2-p_E(t) k_{-2}n_{\rm p}(t)=J_{\rm qst}+J_{\rm geom}.
  \label{nonst} \end{equation}

\section{Geometric rate corrections in a growing cell culture}
The geometric correction (\ref{curcur3}) is generally much smaller
than the main contribution (\ref{curcur2}) if the number of the
enzymes is much smaller than that of the substrates and the products.
However, this small correction has very different properties, and 
can change a system behavior {\em qualitatively} under special
conditions. 

The quasi-steady state contribution to the kinetic
rate in (\ref{curcur2}) can  be vanishing due to a symmetry relation, such as the
detailed balance condition, which guaranties that all chemical fluxes
at the thermodynamic equilibrium state are zero on average. Thus, if a
system is slowly driven externally so that it always remains close to
the thermodynamic equilibrium, the quasi-steady state approximation
will predict zero average product creation. In contrast, the geometric
contribution does not have to remain zero, and it will result in a
qualitatively novel effect.

To show this, consider the MM reaction with concentrations of substrate and product
$n_{\rm s}$ and $n_{\rm p}$ large and treated deterministically. Let
us suppose that the system is initially in an equilibrium, 
\begin{equation}
k_1k_2n_{\rm s}(0)=k_{-1}k_{-2}n_{\rm p}(0).
\label{equil}
\end{equation}
Now suppose that this process happens inside a living cell that grows
and divides
in its usual cycle. Then the substrate/product molecules in a single
cell are diluted by $N(t)$, the number of cells in the descendant colony:
\begin{equation}
n_{\rm s}(t)=n_{\rm s}(0)\frac{N(0)}{N(t)},\,\,\,\,\,\,\, n_{\rm
  p}(t)=n_{\rm p}(0)\frac{N(0)}{N(t)},
\label{decr}
\end{equation}

Since the ratio $n_{\rm s}(t)/n_{\rm p}(t)$ is not affected by this
time dependent dilution, the system remains near equilibrium, and the
quasi-steady state reaction rate remains zero. Thus the average number
of new product molecules, produced by a single enzyme is completely
determined by the geometric part of the rate (\ref{curcur3}),
\begin{multline}
n= \int_{v(0)}^{\infty}dv \left[ -(k_2+k_{-1})
\frac{(k_2+k_{-2}n_{\rm p}v(0)/v)(k_1\partial_{v}({n}_{\rm
    s}(0)v(0)/v)+k_{-2}\partial_v({n}_{\rm p}(0)v(0)/v)}{K^3(v)} \right]=\\
\frac{k_1k_2n_{\rm s}(0)}{(k_2+k_{-1})(k_{-1}+k_1n_{\rm s}(0))}.
\label{deltanp}
\end{multline}

On the one hand, this effect is very small: the average number of new
product molecules per one enzyme is a fraction of unity, which
compares to a large number of already existing substrate and product
molecules. On the other hand, the geometric contribution
qualitatively changes the result, predicting on average nonzero amount
of new product molecules, which is not expected from the standard MM
treatment. If the number of the enzymes in the culture is proportional
to the number of cells, and hence grows with time as $N(t)$, this
effect will eventually become observable.

The result (\ref{deltanp}) would be valid only if we could treat
concentrations as parameters, changing only due to the external volume
growth. In a closed system chemical fluxes eventually should be
compensated by the reverse fluxes due to the violation of the steady
state condition (\ref{equil}). Thus the geometric flux should be
possible to detect by measuring the deviation of the ratio $n_{\rm
  s}/n_{\rm p}$ from the equilibrium value.

Considering intermediate stages of the culture growth, one can notice
that the number of newly produced molecules depends only on the
initial and the final cell numbers: that is, the average number of
produced proteins depends on the current state of the system, but not
on how it got there or where it's going from there. This can be
utilized by living organisms in order to control some processes
depending on the stage of cell's life cycle. Although this effect is
very small, it should be interesting to explore its detectability {\em
  in vivo} and employ it in artificial biochemical circuits design.
 
\section{Discussion}
In this article, we generalized the notion of the geometric phase in
evolution of the mgf to nonperiodic time-dependent processes. For
this, the contour integral of the Berry connection had to be
supplemented by an extra term restoring the gauge invariance of the
geometric contribution to the cgf. This term originates from the
boundary contribution responsible for proper description of the
initial and final moments of the measurement. For nonequilibrium
initial conditions, the boundary terms are responsible for the initial
fast relaxation to the enzymatic quasi-steady state. That is, although
our approach is adiabatic, it also rigorously captures initial fast
relaxation effects.

Our non-periodic geometric phase is different from the ones often
encountered in quantum mechanical applications. Its uniqueness follows
from the existence of a special gauge that should be imposed in order
to describe stochastic kinetics correctly.

We showed that the phase is responsible for nonadiabatic corrections
to the standard Michaelis-Menten approximation. Such corrections are
usually small in comparison to the quasi-steady state predictions.
However, they explicitly break time-reversal symmetries and,
therefore, can produce a qualitatively different result when a
chemical system is driven closely to a thermodynamic equilibrium, as
in the cell culture growth model that we discussed.

It is unclear whether this effect is of biological relevance. However,
we note that we studied only the simplest of its realizations. The
introduced non-periodic geometric phase is completely general and
should appear practically in any interacting chemical system driven by
external fields. Other interesting examples will surely emerge with
time. We expect the greatest opportunities for biological relevance in
the domain of molecular motors, where geometric effects play an
important role as is \cite{astumian-pnas}.

It would also be interesting to understand if the nonperiodic
geometric phase is related to the existence of fluctuation theorems
\cite{jarzynsky-rev}. Indeed, instead of chemical fluxes, it is
possible to use the same formalism to count work or dissipated energy
in a driven stochastic system. The absence of anholonomies, such as
cyclic geometric phases may indicate the existence of fluctuation
relations since then the counting statistics depends only on initial
and final values of external parameters, at least in the adiabatic
limit. Generalizations of our approach to a nonadiabatic evolution
should also be possible since similar generalizations simultaneously
to a noncyclic and nonadiabatic evolution in quantum mechanics exist
\cite{AA}.

\begin{acknowledgments} {N.\ A.\ S. was supported by NSF under Grant No. ECCS-0925618
and partially by the US DOE under Contract No. DE-AC52-06NA25396. I.\ N.\ was supported by Los Alamos National
  Laboratory LDRD program during earlier stages of this work. Authors
  thank Robert Ecke and the entire community of the LANL Center for
  Nonlinear Studies for creation of a unique collaborative research
  atmosphere.} \end{acknowledgments}

\end{document}